*Original Article*

# The Impact of Financial Literacy, Social Capital, and Financial Technology on Financial Inclusion of Indonesian Students

[1]Gen Norman Thomas, [2]Siti Mutiara Ramadhanti Nur, [3]Lely Indriaty

[1,2]Accounting Department, School of Accounting, Bina Nusantara University Indonesia 11480.
[3]Accounting Department, Faculty of Economics and Business, Persada of Indonesia University YAI, Indonesia.

*Abstract: This study aims to analyze the impact of financial literacy, social capital and financial technology on financial inclusion. The research method used a quantitative research method, in which questionnaires were distributed to 100 active students in the economics faculty at 7 private colleges in Tangerang, Indonesia. Based on the results of data processing using SPSS version 23, it results that financial literacy, social capital and financial technology partially have a positive and significant influence on financial inclusion. The results of this study provide input that financial literacy needs to be increased because it is not yet equivalent to financial inclusion, and reducing the gap between financial literacy and financial inclusion is only 2.74%. Another benefit of this research is to give an understanding to students that students should be independent actors or users of financial technology products and that students should become pioneers in delivering financial knowledge, financial behavior and financial attitudes to the wider community.*

## I.INTRODUCTION

Financial inclusion is a problem that needs to be seriously addressed by the government of the Republic of Indonesia in relation to the availability of financial products and services to improve people's economic welfare. Regarding financial inclusion activities, the government, through financial service institutions such as the Financial Services Authority (OJK), has surveyed the level of financial inclusion every 3 years starting in 2013, with the name the Indonesian National Financial Literacy Survey (SNLKI). (SNLKI, 2013) showed a level of financial inclusion of 59.7%. As stated by the Chairman of the OJK Board of Commissioners (Wimboh Santoso, 2022), students are role models who can transmit a culture of fond of saving and investing to the surrounding community and the next generation (beritasatu.com, 2022). Students as agents of development, contribute greatly to accelerating financial inclusion, especially in the banking sector in Indonesia. This research focuses on economics faculty students who are studying financial management because students, as agents of reform, contribute greatly to accelerating financial inclusion, especially in the banking sector in Indonesia. However, students who were expected to make a major contribution to achieving the target of increasing the level of financial inclusion turned out to be that in 2015, there was a very sad case of fraudulent payment of tuition fees by intermediaries using the cash-back mode. A total of 751 certain private college students whom four students previously tricked, were suspected of fraud in paying tuition fees. Hundreds of students pay tuition fees not through official campus mechanisms (kompas.com, 2022). Students who entrust tuition payments through the cash-back system show that the level of student financial inclusion is still low because the cash-back system is not a product of the banking business. Why did fraud occur?

Meanwhile, the results of the 2016 SNLKI produced a score of 67.8%, meaning that there was an increase in financial inclusion of 8.1% for 3 years (SNLKI, 2016). At the 2019 and 2022 SNLKI scores of 76.19% and 85.10% were obtained (SNLKI, 2022). Then, in order to protect the number of victims of fraud, we need legal protection while intensifying efforts to socialize financial inclusion among students.

The government launched the Presidential Regulation of the Republic of Indonesia, Number: 82 of 2016, concerning the National Strategy for Financial Inclusion, which involves institutions to support the program (OECD, 2015). Even though regulations and socialization of financial inclusion have been implemented starting around 2013, cash is very easy to misuse for criminal purposes (Akyuwen and Waskito, 2019), stating that financial inclusion can help society and the economy because financial inclusion can create positive externalities. Positive externality means that the availability of access and benefits to the formal financial system can have a positive influence on all economic actors by providing opportunities to improve welfare if everything goes well.

Another effort made by the government, apart from increasing financial inclusion with financial literacy is to establish an Inclusive Financial Acceleration Village. Tanjung Batu Village is designated as an Inclusive Village by the Financial Services Authority (OJK) of East Kalimantan as part of one of the work programs of the Berau District Regional Financial Access Acceleration Team (TPAKD) to improve inclusive financial policies, in which the community can actively utilize formal financial products and services such as saving to insurance (niaga.asia, 2022). Tanjung Batu Village is designated as an Inclusive Village by the Financial Services Authority (OJK) of East Kalimantan as part of one of the work programs of the Berau District Regional Financial Access Acceleration Team (TPAKD) to improve inclusive financial policies, in which the community can actively utilize formal financial products and services such as saving to insurance (niaga. asia, 2022). The formation of this Financial Inclusion Village can be used as an example of strengthening the aspect of social capital related to one way of increasing financial inclusion. Then, along with advances in information technology, especially advances in the application of financial technology, it certainly has an impact on financial inclusion.

A new phenomenon known as digital financial platforms is emerging as an increasing number of creative businesses enter the market with offerings to reduce the cost and increase the convenience of financial transactions for economic actors (Kabakova & Plaksenkov, 2018). Financial technology has created convenience in using financial services. Financial technology can be accessed anywhere and anytime. Financial services can be easily accessed by smartphones, for example, mobile banking which can be used at any time for transactions within the bank and purchasing financial instruments in the available applications. The Coordinating Minister for the Economy, Darmin Nasution and the Governor of Bank Indonesia (Perry Warjiyo, 2022) agree that financial technology companies are effective in increasing financial inclusion in Indonesia. Because of this, both assess that regulators need to understand the landscape and ecosystem of this industry. According to Darmin (antaranews.com, 2022), financial technology is more effective in encouraging financial inclusion and being able to create a variety of financial products and services. Based on(the OJK Revisit, 2017) conducted by the Financial Services Authority, the use of delivery channels in digital form, namely ATM (Automatic Teller Machine), mobile banking, phone banking and online transactions, has a frequency of use at the top four levels.

Meanwhile, the frequency of using delivery channels per month at financial service offices is at the second lowest level. This shows that people are starting to use and trust digital financial products. Delivery channels in digital form are very easy to access anywhere and anytime so it doesn't take much time to access a product and financial services and requires a relatively low cost. Digital finance has positive and negative (Ozili, 2018). The phenomenon that is happening a lot at the moment is the increasingly uncontrollable new style of fraud against funds that are supposed to be for the benefit of student studies carried out by certain individuals, both from within the campus and from outside the campus, both by means of conventional modus operandi and by means of the use of financial technology. The main problem research gap in this research is the low financial inclusion capabilities of students of the faculty of economics even though students are one of the agents of development. This research aims to find out and analyze the influence of financial literacy, social capital and financial technology on financial literacy. It is still fresh in our memories of the tragedy that occurred in 2015, where almost 1,000 certain private college students were forced to pay tuition fees twice because previously paid tuition fees did not go to the campus treasury but went to fraudsters through the cash-back payment method.

Based on the explanation above, there has been an anomaly in financial inclusion in most students because of the findings of (SNLKI, 2022) that the level of financial inclusion is 85.10% and financial literacy is 49.68%, so there is an increase in financial inclusion of 8.91% and an increase in financial literacy of 11 .65%. How can this happen? Students are a group of intellectual and educated people who are almost impossible to deceive by fraudsters, especially when it comes to economics students. Students who are supposed to be protectors of society have become victims of fraud. How can this anomaly occur? Doesn't financial inclusion continue to increase, but fraud targeting students also increases too. Not to mention, recently, there has been another new style of fraud based on financial technology targeting students, including online loan cases (financial loans) targeting students. It turns out that the online loan fraud case involving the state college student victims in Bogor was in the mode of purchasing fictitious goods, not in the form of online loans (Kontan.co.id, 2022). Based on the fraud incident, it was clearly revealed that the students did not know which one was investing and which one was financing online stores with fictive products. Based on the explanation above, both in cases of fraudulent payment of tuition fees using the cash-back mode and cases of fraudulent investments that turn out to be online shops, financial literacy in students should be corrected. High financial inclusion is less able to be matched by high literacy as well. Other factors that also affect financial inclusion are social capital and financial technology. Why are millennial students who are familiar with financial technology the objects of fraud? The contribution that can be made from this research is to provide valuable input for the government and the world of higher education, especially economics faculties, to start improving themselves to increase financial inclusion by developing academic policies and regulations related to financial inclusion.

## II. LITERATURE REVIEW

This discussion of financial literacy begins with one of the definitions of financial literacy stated by(Lestari, 1970); financial literacy is defined as an awareness or knowledge of financial products, financial institutions and the concept of financial management skills.

Then, the government, through OJK Regulation Number: 76/POJK.07/2016 in article 3 (OJK Regulation, 2016), explains the benefits of financial literacy, including *firs*t, it can improve the quality of decision-making for individuals and *second*, shifts in people's attitudes and behaviors toward money management in order to improve it and enable it to identify and make use of financial institutions, goods, and services that are in line with the needs and capacities of customers and/or the community to accomplish prosperity. After it was felt that there was no explanation about the importance of financial literacy, the government, through the (OJK Institution, 2017), emphasized this again by launching several explanations about the basic principles of financial literacy, namely*: First,* planned and measurable, meaning that the activities carried out must have a concept that is in line with the target, strategies, authority policies and policies of financial service actors**.** *Second*, it has indicators to obtain information related to increasing financial literacy. *Third*, achievement-oriented. Activities carried out are able to achieve the goal of increasing financial literacy by optimizing existing resources, and *Fourth,* Sustainable,

Continuous actions with long-term implications that are taken to accomplish predetermined objectives. Financial service industry players must place a high priority on their knowledge of financial management, institutions, products, and/or financial services when putting the sustainability concept into practice. Social capital is an aspect of the social context that has productive benefits which are realized in an atmosphere of mutual assistance. Social capital certainly produces social benefits (Institute for Social Capital, 2014). Still from the same source, social capital includes solidarity or goodwill between people and groups of people. More than that, in brief, social capital is defined as the behavior of helping each other between people so as to produce feelings of gratitude, mutual respect and friendship. Francis Fukuyama, 2012 describes social capital as a collection of unwritten standards or ideals that members of a group share and adhere to in order to facilitate cooperation. Group members will trust one another if they anticipate that one another will act honorably and consistently. (Fukuyama, 2012). Financial inclusion is the provision of banking services to low-income groups of people at affordable costs (Kaur, 2015). Nowadays, services to the public, especially in the banking sector, are so sophisticated that services are getting faster, all because of information technology. Financial Technology is an industry segment consisting of technology-focused companies with innovative products and services traditionally provided by the financial services industry (Kumail Abbas Rizvi et al., 2018). In plain English, financial technology is the use of technology and financial goods together to make things easier for customers. In actuality, though, the term "financial technology" is often and erroneously used in the business community. The lack of a common definition leads to misconceptions about financial technology that are either overly expansive and boundless or, on the other hand, too restricted to be fully comprehended. (Ilman et al., 2019). Financial technology, according to the Bank of Indonesia, is the application of technology within the financial system that results in new goods, services, technologies, and/or business models and affects the stability of the money supply, the financial system as a whole, and/or the effectiveness, smoothness, security, and dependability of payment systems. According to the Financial Services Authority, Financial Technology is an innovation in the financial services industry that utilizes technology. Financial technology products are usually in the form of a system built to run a specific financial transaction mechanism. (Ozili, 2018) says the term "Fintech" denotes "Financial Technology" defined as the delivery of financial and banking services through modern technological innovations led by computer program algorithms.

*A) Formulation of the Hypothesis*
   **a. The Effect of Financial Literacy on Financial Inclusion**
   In the following discussion, it turns out that according to (OJK, 2017), financial literacy has levels in the following order as follows: ***First and foremost, be well-literate.*** This means that you should be confident in your understanding of financial service organizations and products, including their features, benefits, and hazards, as well as your rights and obligations. You should also be adept at using these goods and services. *Second*, have *sufficient literacy* and confidence in their knowledge of financial service providers, financial goods, and services, including their characteristics, advantages, and hazards, as well as their rights and responsibilities. *Third*, *less literate*, with a limited understanding of financial services, goods, and institutions. *Fourth, lack of literacy*, lack of understanding and trust in financial service providers, financial services and products, and a lack of proficiency in using financial services and products. According to the (OECD, 2018), indicators that can be used to measure financial literacy are, *First*, Financial Knowledge. *Second,* Financial Behavior, and *Third,* Financial Attitudes. This indicator is used to measure the performance of financial literacy on financial inclusion. The results of (Pulungan and Ndruru, 2019) demonstrate how financial inclusion is positively and significantly impacted by financial literacy, with higher financial inclusion being associated with higher financial literacy levels. This is in line with research by (Grohmann, Klühs, and Menkhoff, 2018), which states that there is a positive and significant effect between financial literacy and financial inclusion. Thus, the hypothesis that can be raised is:

*H1: Financial Literacy Affects Financial Inclusion*

**b. The Effect of Social Capital on Financial Inclusion**

According to (Saputra and Dewi, 2017), social capital is solidarity owned, self-confidence and facilities for running a business, which comes from social relations involving family, friends, co-workers and others. Social capital refers to networks, norms and trust to facilitate mutually beneficial cooperation (Okello Candiya Bongomin et al., 2016). There are 6 main elements of social capital (Widodo, 2016), namely: a. Participation in a Network. b. Reciprocity, c. Trust, d. Social Norms, e. Values, and f. Proactive action. Based on (Bongomin et al., 2016), which adopted the World Bank Social Capital Initiative, social capital is measured using dimensions or indicators, namely: 1. Trust, 2. Bonds, 3. Bridging. , and 4. Collective Action. Meanwhile, based on the results of research by (Pulungan and Ndururu, 2019), social capital has a positive and significant effect on financial inclusion because the better the level of trust in financial inclusion, the more development of financial inclusion will increase. This research is also supported by (Safira and Dewi, 2019), who state that there is a significant relationship between social capital and financial inclusion. Based on the description above, the hypothesis that can be raised is:

*H2: Social Capital Affects Financial Inclusion*

**c. The Effect of Financial Technology on Financial Inclusion**

Based on the (OJK, 2022), several types of Fintech are currently developing and providing financial solutions for Indonesian people, namely: ***Initially, Crowdfunding:*** A financial technology approach that is growing in popularity in many nations, including Indonesia, is crowdfunding. People can use this technology to generate money for social programs or initiatives that are important to them. Raising money to construct the R80 aircraft, which was designed by BJ Habibie, is one instance. An illustration of a finance startup that is now well-liked in Indonesia that uses a crowdfunding approach is (Kita Bisa.com, 2022). ***Second is Microlending***: One fintech service that helps members of the lower middle class with their everyday lives and finances is microfinancing. The majority of persons in this economic category lack access to banking institutions, and they also have trouble getting venture capital to grow their companies or means of subsistence. By directly connecting potential borrowers with lenders of business capital, microfinancing aims to address this issue. The structure of the business system is such that profits are attractive to lenders but within reach of borrowers. Amartha is one of the startups involved in microfinance; it links online investors with rural micro-entrepreneurs. ***Third. P2P Lending Service:*** Fintech is a more popular term for this kind of lending. This technology assists people who require access to financing to suit their demands. Customers can borrow money more readily with this Fintech to satisfy a variety of life's necessities without having to go through the complicated procedures that are frequently present in traditional banks. One fintech company operating in the money lending space is AwanTunai, a startup that offers convenient and safe digital installment options. ***Fourth. Market Comparison***: Users can compare different financial products from different financial service suppliers with this technology. Financial planning is another role that Fintech may play.

Fintech provides people with a variety of investing possibilities to suit their demands going forward. ***Fifth. Digital Payment System***: this Fintech company offers services for paying all bills, including credit and postpaid ones, credit cards, and tokens for PLN electricity. One fintech company that uses a digital payment method is Payfazz. This agency-based business assists Indonesians, particularly those without bank access, in making monthly payments for bills of every kind. Based on (Ozili, 2018), digital finance through financial technology providers has a positive effect on financial inclusion. This result is also in line with the research of (Kabakova and Plaksenkov, 2018), which states that digital progress is accompanied by a suitable economic environment and a strong socio-demographic component of the ecosystem leads to higher financial inclusion. Based on the explanation above, the hypothesis that can be raised is:

*H3: Financial Technology affects Financial Inclusion*

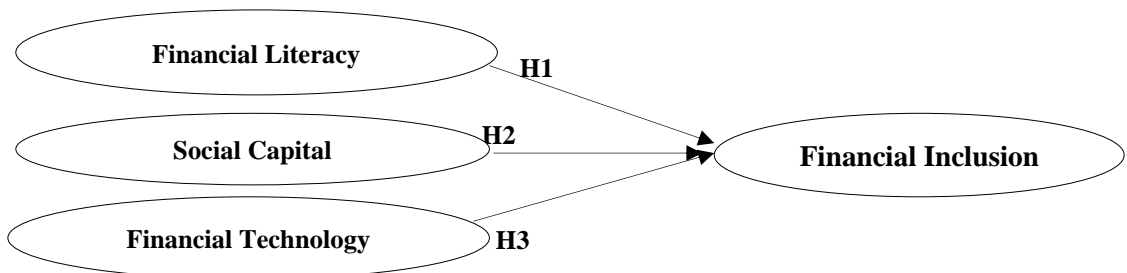

**Figure 1: Research Hypothesis**

*B) Research Methods*

The research method used was the quantitative research method. Three research hypotheses were tested with research objects from economics faculty students who were actively carrying out their studies as primary data. Students are included as targets for increasing financial inclusion, which is influenced by financial literacy, social capital and financial technology. Primary data is obtained and processed to then take conclusions. (Sugiyono, 2018).

The objects of the study were 5th-semester students of the economics faculty of management, business and accounting programs who are domiciled in Tangerang, Indonesia, who are currently studying at 7 universities, namely Bina Nusantara University (UBN), Gunadarma University (UG), Matana University (UM), Muhammadiyah University Tangerang (UMN) Multimedia Nusantara University (UMN), Pelita Harapan University (UPH), and Swiss German University (USG). Based on a population of less than 1,000 students, 10% of the 1,000 students were determined, so 100 respondents were selected (Arikunto, 2012). Primary data collection was carried out by submitting questionnaires directly (Sugiyono, 2018), and data processing was carried out using the SPSS version 23.

**Table 1: Operationalization of Variables**

| Variables | Indicators | Sources |
|---|---|---|
| Financial Literacy (X1) | Financial Knowledge<br>Financial Behavior<br>Financial Attitudes | OECD, 2018 |
| Social Capital (X2) | Collection Action<br>Bonding<br>Bridging<br>Trust | Bongomin, et,al, 2016 |
| Financial Technology (X3) | Financial Technology | Bank Indonesia, 2019 |
| Financial Inclusion (Y) | Access<br>Usage<br>Quality | Bank Indonesia, 2019 |

*Sources: Data processed by SPSS, version 23, 2022*

## III. RESULTS AND DISCUSSIONS

*A) Results*

**a. Identity of Respondents**

The results showed that 100 respondents who received the questionnaire had answered all questions completely, and there were no errors in filling out; with details from UBN, there were 51 respondents, UG had 18 respondents, UM had 1 respondent, UMT had 10 respondents, UMN had 12 respondents, UPH there were 7 and USG there was 1 respondent, Respondents based on gender, 36% male and 64% female. Meanwhile, respondents based on income can be seen in the following categories: income below IDR 2,000,000, as much as 44%, income IDR 2,000,000 to IDR. 4,000,000, as much as 35%, while income above IDR 4,000,000 is as much as 21%.

**b. Validity and Reliability Test**

Based on the results of the validity test, it is known that the rcount of each indicator on financial literacy, social capital, financial technology and financial inclusion is all greater than the rtable with a degree of freedom of 98 (100-2) in SPSS version 23, the results show that all questions research questions declared valid for primary data collection for this study. Meanwhile, the results of the reliability test on the indicators for all variables show the results of Cronbach's Alpha test > 0.700, which means that all the questions asked in the questionnaire are declared reliable. After the quality of the questions is declared valid and reliable, the next test is the classical assumption test.

**Table 2: Descriptive Statistics**

| Variables | N | Min. | Max. | Sum | Mean | Standard Deviation |
|---|---|---|---|---|---|---|
| Financial Literacy | 100 | 55 | 75 | 6,370 | 63,70 | 5,734 |
| Social Capital | 100 | 27 | 40 | 3,361 | 33,61 | 3,547 |
| Financial Technology | 100 | 15 | 40 | 3,222 | 32,22 | 5,486 |
| Financial Inclusion | 100 | 45 | 70 | 5,757 | 57,57 | 6,458 |
| Valid N (listwise) | 100 | | | | | |

*Sources: Data processed by SPSS, version 23, 2022*

The normality test using the Kolgomorov-Smirnov test produces Asymp. Sig. (2-tailed) > 0.05, which is 0.200. Based on these results, it can be concluded that the data in this study were normally distributed. In the same test using a histogram chart, it is known that the data is also normally distributed because the graph is symmetrical and bell-shaped. The results of the multicollinearity test can be seen in the tolerance value and the Variance Inflationary Factor (VIF) value. Financial literacy (X1) has a tolerance value of > 0.10, which is 0.678 and a VIF value of < 10, which is 1.475, while social capital (X2) has a tolerance value of > 0.10, which is 0.704 and a VIF value < 10, which is 1.420.

Finally, financial technology (X3) has a tolerance value of > 0.10, which is 0.691 and a VIF value of < 10, which is 1.447, so it can be concluded that financial literacy (X1), social capital (X2) and financial technology (X3) in this study are not multicollinearity occurs. In the heteroscedasticity test, financial literacy (X1) has a significance value > 0.05, namely 0.233. social capital (X2) has a significance value of > 0.05, namely 0.976. In financial technology (X3), it has a significance value of > 0.05, namely 0.088. It can be concluded that all the independent variables in this study did not occur heteroscedasticity.

**Table 3: Multiple Regression Analysis**

| Model | | Unstandardized Coefficients | | Standardized Coefficients | t | Sig. |
|---|---|---|---|---|---|---|
| | | B | Std Error | Beta | | |
| 1 | (Constant) | 7,305 | 4,907 | | 1,489 | ,140 |
| | Financial Literacy | .183 | .085 | .162 | 2.139 | .035 |
| | Social Capital | .642 | .136 | .353 | 4.740 | .000 |
| | Financial Technology | .529 | .088 | .449 | 5.974 | .000 |
| | | | | | | |

a. Dependent variable: Financial Inclusion

*Sources: Data processed by SPSS, version 23, 2022*

Based on the table above, it can be seen that the Unstandardized Coefficients column in B shows a constant value (□) of 7.305, and the next row shows the coefficient of the independent variable. The following is a multiple linear regression equation based on the results of data processing.

$$Y = 7{,}305 + 0{,}183\ X_1 + 0{,}642\ X_2 + 0{,}529\ X_3 + \varepsilon$$

Table 3 shows that financial literacy has a t count > t table, namely 2.139 > 1.985 and a significance value < 0.05, namely 0.035 < 0.05. This shows that Ha is accepted; namely, the financial literacy variable (X1) has a significant influence on financial inclusion (Y), social capital (X2) has, t count > t table, namely 4.740 > 1.985 and a significance value <0.05, namely 0.000 < 0.05 also shows that Ha is accepted, namely social capital (X2) has a significant influence on financial inclusion (Y), and financial technology (X3) has a t count > t table, namely 5.947 > 1.985 and a significance value < 0.05, namely 0.035 < 0.0 5 thus indicating that Ha is accepted, namely the financial technology variable (X3) has a significant influence on financial inclusion (Y).

**Table 4: Determination Coefficient Test**

| | | Model Summary[b] | | |
|---|---|---|---|---|
| Model | R | R Square | Adjusted R | Std Error of the Estimate |
| 1 | .791[a] | .625 | .614 | 4,014 |

*Sources: Data processed by SPSS, version 23, 2022*

Based on Table 4, the Adjusted $R^2$ value is 0.614 or 61.4%. This shows that the independent variables, namely financial literacy, social capital and financial technology, have a relationship with the dependent variable, namely financial inclusion of 61.4%. Meanwhile, the remaining 0.386 or 38.6% is influenced by variables or other factors not included in the study.

c. **The Effect of Financial Literacy on Financial Inclusion**

Based on the results of the hypothesis, it shows that financial literacy affects financial inclusion with a t count > t table, namely 2.139 > 1.985 and a significance level of 0.035. Therefore, it can be concluded that H1 is accepted; namely financial literacy has a positive and significant effect on financial inclusion in economics students in the Tangerang area. It can be concluded that the higher a person's financial literacy in terms of financial knowledge, financial behavior and financial attitudes, the higher the level of financial inclusion. A person with high financial literacy will be able to make appropriate decisions about the products and services they use, plan their finances more effectively, steer clear of investing in dubious financial instruments, and comprehend the advantages and disadvantages of various financial offerings. (Lestari, 2015).

This is in line with the research of (Pulungan and Ndruru, 2019); financial literacy has a positive and significant effect on financial inclusion, so the better the level of financial literacy, the higher financial inclusion. This is also in line with research(Grohmann et al., 2018), which states that there are positive and significant results in the relationship between financial literacy and financial inclusion. But don't forget that the effect can also be the opposite; that is, the lower the level of financial literacy, the lower the financial inclusion that can be utilized. This effect is also positive.

**d. The Effect of Social Capital on Financial Inclusion**

Based on the results of the hypothesis, it shows that social capital has an effect on financial inclusion with a t count > t table, namely 4.740 > 1.985 and a significance level of 0.000. Therefore, it can be concluded that H2 is accepted, namely social capital has a positive and significant effect on financial inclusion in undergraduate students in the Tangerang area. It can be concluded that the higher a person's social capital, the higher the level of financial inclusion. Social capital is crucial in promoting resource sharing, which includes the neighbourhood's expertise and abilities, which are the main drivers of financial literacy (Saputra & Dewi, 2017). Social capital through trust, bonding, bridging and collective action can increase the level of financial inclusion of Tangerang students. Through the role of social capital, it can channel and create trust in the use of financial services and services, so that the level of financial inclusion can increase. This is in line with the results of research by (Pulungan and Ndruru, 2019), which proves that social capital has a positive and significant effect on financial inclusion because the better the level of trust in financial inclusion, the more development of financial inclusion will increase. This is also in line with the research (Safira & Dewi, 2019), which states that social capital has a significant effect on financial inclusion. But don't forget that this effect can also be positive in reverse, namely, the lower the level of social capital, the lower the financial inclusion that can be utilized.

**e. The Effect of Financial Technology on Financial Inclusion**

Based on the results of the hypothesis, it shows that financial technology has an effect on financial inclusion with a t count > t table, namely 5.947 > 1.985 and a significance level of 0.000. Therefore, it can be concluded that H3 is accepted, namely that financial technology has a positive and significant effect on financial inclusion for undergraduate students in the Tangerang area. It can be concluded that the higher a person's financial technology, the higher the level of financial inclusion. With the existence of financial technology, it is easy to access and use financial services that exist on financial technology platforms, namely digital payments, crowdfunding, fintech funding, peer-to-peer lending and market comparison can be done anywhere, so that someone's access to financial services is even easier formal. This is in line with (Ozili, 2018) that digital finance through financial technology providers has a positive effect on financial inclusion. This is also in line with the research (Kabakova & Plaksenkov, 2018) states that a suitable economic environment accompanies digital progress, and a strong socio-demographic component of the ecosystem leads to higher financial inclusion. But some things are very important that also have a positive effect in parallel, namely the lower the level **of** financial technology, the lower the financial inclusion that can be utilized.

*B) Discussions*

Based on the findings of this study, financial literacy, social capital, and financial technology partially have a positive and significant effect on financial inclusion; however, there are still a number of issues that need to be discussed. ***First,*** on the aspect of the effect of financial literacy on financial inclusion and in accordance with the findings of the 2019 and 2022 SNLKI, the gap between financial inclusion and financial literacy in 2019 was 38.16% (76.19% - 38.03%), while the same gap in 2022 was 35.42% (85.10% - 49.68%). A good gap is a gap that is getting smaller, meaning that the level of financial literacy approaches the level of financial inclusion until it is balanced. Taking into account the difference in the gap between 2019 and 2022, there is actually no significant reduction in the gap between financial inclusion and financial literacy. Students of the Faculty of Economics, although they study economics and finance every day, still do not fully understand the types and benefits or products of banking, so they are not able to take full advantage of the use of banking products and services. In other words, the availability of products, services and financial benefits is available before the knowledge, confidence and ability of students to use them; as a result, several products, services and financial benefits have not been utilized. This condition is an easy target for criminals to deceive students. ***Second,*** from the aspect of the influence of social capital, starting from collective action, bridging, trust to attitudes have a positive and significant effect on financial inclusion. Student financial inclusion has not fully experienced an increase influenced by social capital because students, as we know, are not fully capable and willing to live a social life with the community. Besides because, students do not yet have adequate knowledge of finance and banking services.

***Finally,*** the financial technology aspect has a positive and significant influence on financial inclusion. Characteristics of students are usually lazy to think and often choose the easiest way to do something, including taking care of their own interests.

Students avoid using financial technology which is a bit difficult, even though they are actually avid game players. So, it is not surprising that various new modes of fraud based on financial technology will reappear in the future as long as there has been no concrete progress in financial inclusion.

## IV. CONCLUSION AND SUGGESTION

The effect of financial literacy, social capital and financial technology on financial inclusion partially has a positive and significant influence. However, lately, there have been many cases of fraud related to financial technology, which has made students the targets of fraud. Based on the research results, financial inclusion is very good and complete. It has been able to be balanced by various types of all-digital service facilities based on financial technology. The limitation of this research is on students as research objects because it understands the conditions of students whose activities are not yet related to financial inclusion. Students do not have their own income and financial activities are also limited. There are several things that need to be suggested, namely: first, students must study in depth the risks that occur if they are too bold to hand over financial management problems to other parties for any reason.

Understanding the characteristics of a banking product, service, method of payment and all other financial affairs with all the advantages, disadvantages and convenience with all the risks is an important element for increasing financial literacy and inclusion. Second, don't trust people who work for our interests too easily but still believe that the system resolves all matters. Students must be actors, not just spectators.